\documentclass[reprint,rmp,amsmath,amssymb,aps,floatfix]{revtex4-2}

\usepackage{graphicx}
\usepackage{dcolumn}
\usepackage{bm}
\usepackage{subcaption}
\usepackage[font={small,rm},labelfont=bf,textfont={it}]{caption}
\usepackage{upgreek}
 \usepackage{lineno}
\renewcommand{\arraystretch}{1.5}
\newcommand\numberthis{\addtocounter{equation}{1}\tag{\theequation}}
\newcommand{\rom}[1]{\text{\uppercase\expandafter{\romannumeral #1\relax}}}

\newcommand{\smalldpi}{5}
\newcommand{\includegraphicsdpi}[3]{
    \pdfimageresolution=#1  % Change the dpi of images
    \includegraphics[#2]{#3}
    \pdfimageresolution=72  % Change it back to the default
}

\usepackage[%  
    colorlinks=true,
    pdfborder={0 0 0},
    linkcolor=red,
    citecolor=cyan
]{hyperref}
\begin{document}
\preprint{APS/123-QED}

\title{Boundary dynamics in competing critical black hole formation} 

\author{Cole Kelson-Packer}
\email{ckelsonpacker99@unm.edu}
\affiliation{Department of Physics and Astronomy, University of New Mexico\\
  Albuquerque, New Mexico 88003, USA}
\author{John Belz}
\email{belz@physics.utah.edu}
\affiliation{Department of Physics and Astronomy, University of Utah\\
Salt Lake City, Utah 84112, USA}
\date{\today}

\begin{abstract}
Expanding upon our previous study of competing critical phenomena in black hole formation, we numerically investigate the behavior of dominant exponents across the boundary separating asymptotically dispersing and collapsing regions in a two-dimensional configuration space of initial data. We find that across the Type~II boundary section the dominant exponent remains constant, equal to the reciprocal of Choptuik's well-known quasi-universal value, whereas across the Type~I section the exponent noticeably varies. We postulate that this change reflects the existence of a third critical solution in addition to the two primary competing solutions, possibly another member of the family of metastable soliton stars constituting the Type~I attractor.
\end{abstract}
\maketitle

\section{Introduction}
\label{sec:intro}
\par The study of black hole genesis has yielded a wealth of insights into the behavior of classical gravity in the strongly interacting regime. One of the cornerstones of numerical relativity is the existence of critical phenomena accompanying their formation~\cite{choporiginal}. Generically, it has been found that when some parameter (e.g. amplitude) characterizing the abundance of matter sources in the initial data is varied there exists a sharp delineation between the sources' collapse into a black hole and asymptotic dispersal~\cite{c1,c2,c3}. Time evolutions of initial data near this threshold exhibit various quasi-universal properties~\cite{choporiginal}, such as mass power laws with particular exponents, (discrete) self-similarity in the form of echoing effects, and predictable scaling in the time-to-collapse of the lapse~\cite{gundlachrev,gundreview}. It is by the analogy of these with the physics of near conventional critical points in statistical mechanics that this behavior is referred to as critical phenomena.
\par In our previous paper~\cite{us} we observed the apparently counterintuitive effects of multicritical collapse, whereby two near-critical fields appeared to frustrate, rather than accentuate, the process of collapse. The starting point for our investigation was the idea that some kind of interesting interaction should be observed between near-critical massive and massless scalar fields. Since the massive scalar field exhibits Type~I and Type~II criticality in different regions of the parameter space of initial data~\cite{othermassive}, it might be reasonably hypothesized that such configurations are especially susceptible to Type~II perturbations in the Type~I phase. We observed that the two fields, when both are tuned near criticality, appear to inhibit each other's collapsing tendencies, and suggested that this result may be attributed to the existence of a third critical solution, similar to a scenario proposed by Gundlach et al.~\cite{panic}. 
\par
\begin{figure*}[!htp]
\centering
\includegraphicsdpi{\smalldpi}{width=\linewidth}{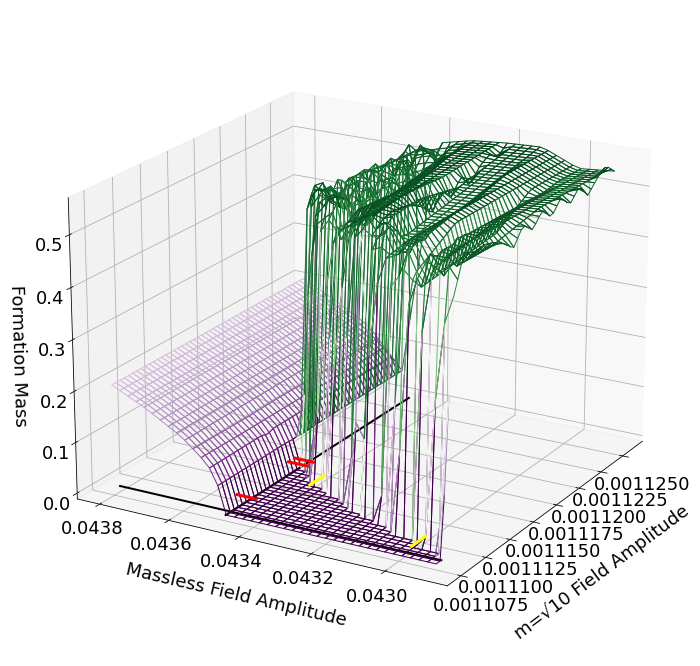}
   \caption{\label{fig:combinedcolor} Phase diagram of evolution behavior for the multicritical field configuration considered in this paper and its predecessor. The x and y variables correspond to the amplitudes tuning the initial data of the scalar fields that ultimately determine whether the time evolution of that data collapses or disperses. The z variable reflects the mass of the black hole formed, with asymptotic dispersal signified by a mass of zero. The black lines reflect the amplitudes greater than which, if one field were taken alone, a black hole would form, approximately $\sim 0.001111$ for the massive field and $\sim 0.04347$ for the massless.  We repeat the observation that the above picture suggests the presence of a mechanism qualitatively inhibiting black hole formation.}
\end{figure*}

\begin{figure}[!ht]
  \centering
  \includegraphicsdpi{\smalldpi}{width=0.9\linewidth}{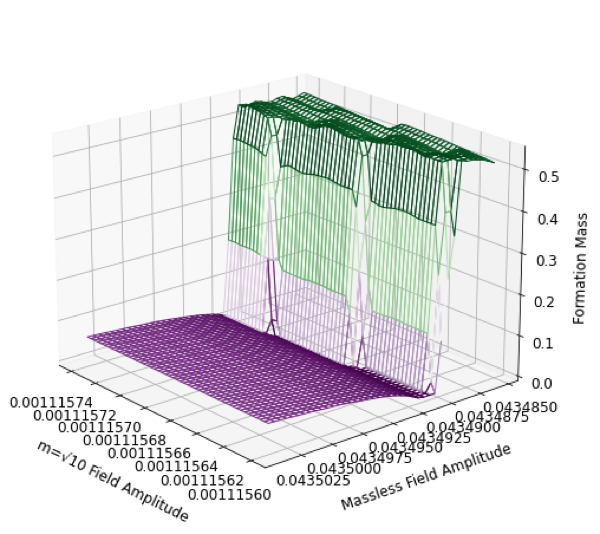}
   \caption{\label{fig:meshzoomrefine} Mesh plot of a refined region about the triple point. The data here comes from jobs ran at high resolutions, showing that the oscillations portrayed in the Type~I plateau are real and illustrating the sharpness of the intersection of the two sections of the boundary. As this is close to the triple point, the dispersal region has been reduced to a mere sliver.}
\end{figure}

\begin{figure}[!htp]
  \centering
\includegraphicsdpi{\smalldpi}{width=0.9\linewidth}{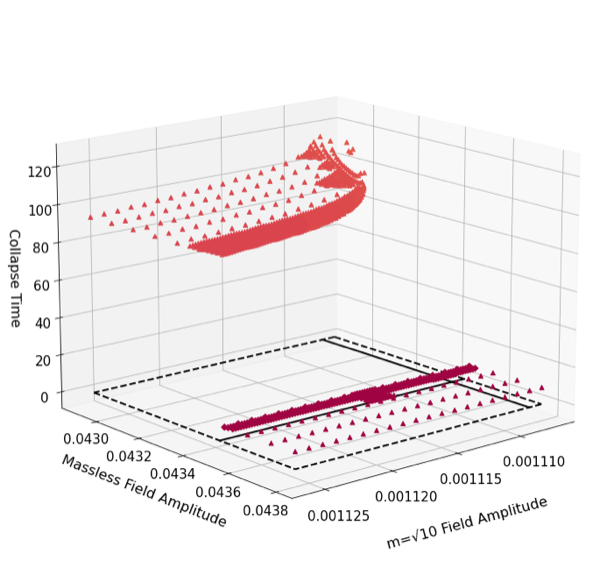}
 \caption{\label{fig:combinedcolortime} Another phase diagram similar to Fig.~\ref{fig:combinedcolor} above, here expressed in terms of asymptotic time to the collapse of the lapse. Triangles indicate Type~I collapse while diamonds denote Type~II. The dashed square at the bottom circumscribes sampled data; comparing with Fig.~\ref{fig:combinedcolor}, scenarios with zero mass there do not appear here, since there is no collapse. While Type~I points curve noticeably upwards near the dispersal boundary, this effect does not manifest across the Type~II portion -- all as expected.}
  %The circles indicate no collapse, the triangles denote Type~I collapse, and the diamonds are Type~II collapse. In the dispersing region the collapse time has been set to $t_{collapse}=800$. Three distinct plateaus appear, consistent with the three qualitatively different regions in Fig.~\ref{fig:combinedcolor}.}
\end{figure}

 Figure~\ref{fig:combinedcolor} reflects the results of our first paper, here cast into a wire mesh plot. It depicts a phase diagram with respect to initial amplitudes for a massless and a massive scalar field minimally coupled to gravity. The z-axis reflects the black hole formation mass derived from the radius at the asymptotic time of collapse, given the initial coordinate amplitudes. The lowest values at zero mass (dark purple online) indicate asymptotic dispersal. The thick lines in the amplitude plane denote the critical amplitudes greater than which colapse would occur if only a corresponding field was present, approximately  $0.0435$ for the massless field and $0.00111$ for the massive. We additionally show in Fig.~\ref{fig:meshzoomrefine} the region about the triple point in greater detail, as well as a 3d scatter plot of the time to collapse in Fig.~\ref{fig:combinedcolortime}. The two plateaus in this last plot correspond to the distinct critical behaviors seen in Fig.~\ref{fig:combinedcolor} through their noticably different times-to-collapse.
\vfill\null
\par While the initial waveforms were chosen to accentuate the effects, the fact that dispersal scenarios exist at larger amplitudes even past the intersection of the critical lines of Fig.~\ref{fig:combinedcolor} -- that is, that the coordinate values of the ``triple point'' are greater than the two critical amplitudes -- suggests that the two fields have a mutually inhibitory effect on each other. This is a nonintuitive effect: what is in a naive sense a greater concentration of mass-energy has the result of interfering, rather than augmenting, collapse.
\par We suggested in our previous work that a rough dynamical systems explanation suffices to explain our results. The intuition is that the earlier critical evolution of the massless field in a sense draws the the spacetime away from the critical surface corresponding to the massive field that would otherwise determine asymptotic behavior. A simple interpretation might reduce this to a triviality in terms of an exchange of energy between the two fields, wherein the massless field implodes through the origin, carrying away some of the energy of the massive field. What is significant is that this physical argument could just as well be turned the other way: that is, the concentration of the massive field about the origin could be suspected to focus and retain the massless field. The dynamical systems sense, however, gives us the argument that the (locally in time) dominant exponent dictates the actual course of evolution, which implicates the frustration we observe. This also explains why the dispersal region impinges so considerably across the Type~I critical vertical line -- yet barely across the Type~II critical horizontal line -- in Fig~\ref{fig:combinedcolor}.
\par The purpose of this paper is to more rigorously quantify the dynamical mechanisms at work so as to solidify assertions made in our previous paper, as well as explain other phenomena we have found since. This quantification justifies the loose time-to-collapse classification scheme used previously, showing how very different perturbations predominate across different parts of the collapse/dispersal boundary. We also discuss how an observed change in the value of the exponent associated with the dominant perturbation along the Type~I section of the boundary suggests richer dynamical phenomena involving the Type~I solution, possibly attributable to a third critical solution. These effects may explain additional numerical complexities we have found along the Type~I section of the boundary further from the triple point.

\section{Methods}
\label{sec:methods}
\par The methods employed in our numerical simulations are the same standard techniques~\cite{baumpiro,alucart,obligatory} as those adumbrated in our previous paper~\cite{us}. We summarize our methodology in the following.
\par We take the standard ADM decomposition assuming spherical symmetry and work in the polar areal gauge, corresponding to the line element $ds^2=-\alpha^2dt^2+a^2dr^2+r^2d\Omega^2$~\cite{ADM}. Matter in the form of massless and massive scalar fields is evolved in time via step-doubling fourth-order Runge-Kutta using the coupled equations
\begin{eqnarray}
\partial_t \Phi_i & = & \frac{\alpha}{a}\Pi_i, \hspace{4.2cm} i = 1,2, \nonumber \\
\nonumber \\
\partial_t \Psi_i & = & \partial_r \left( \frac{\alpha}{a} \Pi_i \right), \hspace{3.3cm} i = 1,2, \nonumber \\
\partial_t \Pi_i & = & \frac{1}{r^2} \partial_r \left( \frac{\alpha r^2}{a} \Psi_i \right) - \alpha a m^2_i \Phi_i, \hspace{0.85cm} i = 1,2, \nonumber \\
 & & \hspace{4.8cm} m_2 = 0. 
\end{eqnarray}
\par Simple Sommerfeld boundary conditions are imposed at large $r$, and appropriate anti-/symmetry constraints are taken across a staggered origin. Meanwhile, the metric equations are solved on each full time step (extrapolated during intermediate steps) following
\begin{equation} \label{eq:aevolve}
\partial _ra = \frac{a}{2}\left[ \frac{1-a^2}{r}+\frac{r}{2} \sum _{i=1}^{2} (\Pi _i^2+\Psi _i^2)+\frac{m_1^2r}{2}a^2\Phi _1^2\right],
\end{equation}
\begin{equation} \label{eq:alevolve}
\partial _r\alpha = \alpha \left[ \frac{\partial _ra}{a}+\frac{a^2-1}{r}-\frac{m_1^2r}{2}a^2\Phi _1^2 \right],
\end{equation}
subject to the boundary conditions of $a$ being unity at the origin and $\alpha$ taken asymptotically Schwarzschild-like. Finally, in our previous paper we verified convergence against the momentum constraint 
\begin{equation}
0=M\equiv \alpha \tfrac{r}{2}(\Pi _1\Psi _1+\Pi _2\Psi _2)-\partial _ta,
\end{equation}
which validates our code following typical tests, showing it to be fourth-order accurate as designed. Finally, to resolve near-critical Type~II behavior about the origin we use familiar methods employing multiple spatial grids, refining by a factor of two or four on each level up to five subgrids near the origin. Refinement and unrefinement are conditioned on thresholds of momentum constraint violation. Most data comes from simulations employing $64000-256000$ coarse radial gridpoints, with the radial grid extending out up to $r=800$ in analysis of Type~I cases and truncated by an order of magnitude for the most precise Type~II cases.
\par This paper employs standard elementary methods for numerically analyzing dynamical systems~\cite{stuarthumphries,strogatz,teschl}. Near criticality, we expect that appropriately dimensionalized functions $Z_p(x,t)$ may be expanded as
\begin{equation} \label{eq:perturb}
Z_p(x,t) \approx Z^{*}(x,t) + C(x)(p-p^*)e^{\gamma t} + ...\mbox{  } ,
\end{equation}
where $Z*$ denotes the function $Z_p$ when the tuning parameter $p$ is at criticality $p^*$, $\upgamma$ is the most dominant perturbative exponent, and $t$ is a time variable appropriate to the critical system~\cite{gundlachrev,gundreview}. When analyzing Type~I scenarios, we take $t$ to be the difference between the asymptotic time corresponding to the space-like slice of the coarsest grid and the time to collapse in the same measure, both obtained from integration of the lapse at the radial edge of the simulation. In the Type~II case, we consider the negative logarithm of the negative of this quantity instead. We shall refer to the quantity $\upgamma$ derived in these two cases as $\upgamma_{\rom{1}}$ and $\upgamma_{\rom{2}}$ respectively.
\par For the purposes of presentation, we consistently take $p$ to be the amplitude $A$ of an initially ingoing spherically-symmetric Gaussian shell with profile
\begin{equation}
  \phi(r,0)=Ar^2\exp(-(r-r_0)^2/\sigma^2),
\end{equation}
where $\sigma=1.0$ or $5.0$ for the massless and massive fields respectively, and $r_0=2.0$ for both fields. This configuration was specifically chosen such that the massive field undergoes Type~I evolution and that both fields interact substantially early on, thus magnifying the dynamical effects of competition. By virtue of the mass, the two fields will arrive at the origin at different times; numerical considerations involving the growth of the amplitudes of ingoing waves have prevented us from arranging simultaneity while maintaining respectable precision near criticality in certain parts of the parameter space.
\par We variably take the perturbative $Z$ to be $\Phi$, $r\Pi$, or $r\Psi$. For all simulations performed, each choice leads to the same essential results and conclusions.
%The specific choice is purely aesthetic, in that for all simulations performed each choice lends itself to the same essential results and conclusions. Not all initial conditions will lead to the same competition observed, in the sense that the ``dispersal region'' isn't anomolously large, but most observations apply mutatis mutandis.
\vfill\null
\par We perform this analysis by taking the difference of two data sets very close to criticality, taking the $L^2$ norm of this quantity, and then carrying out an appropriate regression to obtain an estimate for $\upgamma$. Explicitly, we take the difference of Eq.~(\ref{eq:perturb}) for two distinct values of $p$,
\begin{equation}
Z_{p_1}(x,t)-Z_{p_2}(x,t) \approx C(x)(p_1-p_2)e^{\gamma t} + ...\mbox{  },
\end{equation}
then take the $L^2$ norm to remove spatial dependence, resulting in the approximation
\begin{equation}
y(t) \approx C(p_1-p_2)e^{\gamma t} +...\mbox{  }.
\end{equation}
\par This is in a form readily subjected to regression analysis.
\par This regression is typically valid for a reasonable time interval near collapse. We determine collapse by monitoring when the lapse drops beneath $10^{-7}$, or when NaN errors arise from the substantial curvature developing at the origin. The adjective ``reasonable'' cannot be dispensed: data taken too close to collapse is naturally subject to nonlinearities, with the higher order terms becoming relevant, while data coming from too long before features significant contributions from subdominant terms. Since fine-structure undulations appear in these regressions as well, the uncertainties for the derived exponents are greater than those suggested by the regressions alone. %Given also that fine-structure undulations inevitably feature in these regressions as well, conducting these regressions is something closer to an art than a rigorous science, which is to say that the logical margin of error for the $\gamma$s are greater than those suggested by the regressions alone.
\par We check the validity of these perturbative regression estimates against other observable quantities dependent upon $\upgamma_{\rom{1}/\rom{2}}$ and the tuning parameter $p$'s deviation from its critical value $p^*$. For Type~II collapse, we compare with a regression on a local measure of black hole initial formation mass near criticality~\cite{gundconst1}:
\begin{equation}
M \approx A(p-p^*)^{1/\gamma_{\rom{2}}} ,
\end{equation}
where, for our purposes, we will be ignoring the well-known fine-structure corrections~\cite{eldrant}, although their undulations are readily observed in the results \textit{infra}. Meanwhile, for Type~I collapse, we cross analyze with the collapse time:
\begin{equation}
T_{collapse} \approx \rm{const}. - \frac{1}{\gamma_{\rom{1}}} \log(|p-p^*|),
\end{equation}
which, like above, we obtain from integration of the lapse at the edge of the simulation.
\par Numerically solving for and perturbing the exact solution would provide much more precise values of the dominant exponent $\upgamma$~\cite{gundconst2}. However, this would not seem so feasible for our scenario, which admits features that might preclude the simplest implementations of both the single-variable discrete self-similarity of the Type~II exact solution~\cite{gundconst1}, as well as the simple single metastable soliton star solution we obtain in the purely massive field Type~I case~\cite{othermassive,suen,suen2}. We obtain regardless sufficient agreement between the perturbative analysis and criticality probes to make this more exacting precision unnecessary for the particular results we report and our analysis thereof.

\section{Results}
\subsection{Type~II Boundary}
\par We find that the massive field does not significantly affect the fundamental dynamics of Type~II criticality. Quantitatively, we find that although the presence of a massive scalar field does have the inhibitory effect of slightly raising the massless critical parameter (from $\approx 0.043479$ to $\approx 0.043492$), it does not alter the dominant exponent associated with the Choptuon in the limit of the pure massless field.
%\par Figures \ref{fig:typeIIregs} and \ref{fig:typeIIperturbs} illustrate our point. Each panel in Fig.~\ref{fig:typeIIregs} depicts a linear regression taken with respect to the logarithm of the masses arising from black hole formation near the Type~II/dispersal boundary. The massive field amplitude is held fixed, while the logarithm of the deviation of the massless amplitude from criticality is reflected in the abscissa. Figure \ref{fig:typeIImergereg} depicts all these regressions in a single figure along with the pure massless case for ready comparison by eye. Meanwhile, each panel in Fig.~\ref{fig:typeIIperturbs} reflects a perturbative analysis with data taken near criticality for the massless field. Again, the massive is held at fixed amplitude, with the value equaling that fixed for the panel in Fig.~\ref{fig:typeIIregs} commensurate in position.
\par
\begin{figure*}[!ht]
\centering
\begin{subfigure}{0.33\linewidth}
    \subcaption{\large{Massive Field = 0.0006}}
\includegraphicsdpi{\smalldpi}{width=\linewidth}{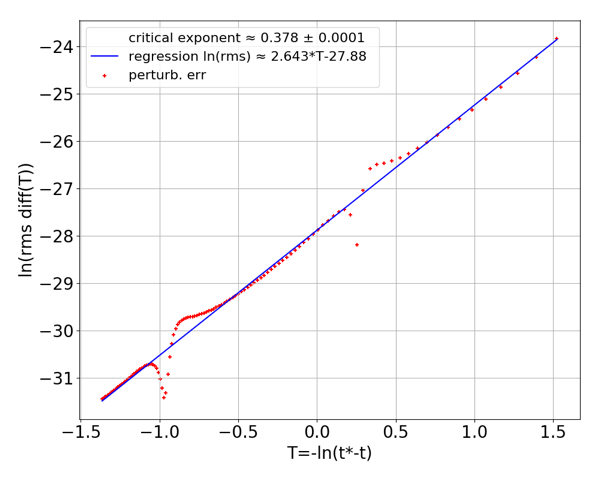}
\end{subfigure}%
\begin{subfigure}{0.33\linewidth}
    \subcaption{\large{Massive Field = 0.0008}}
\includegraphicsdpi{\smalldpi}{width=\linewidth}{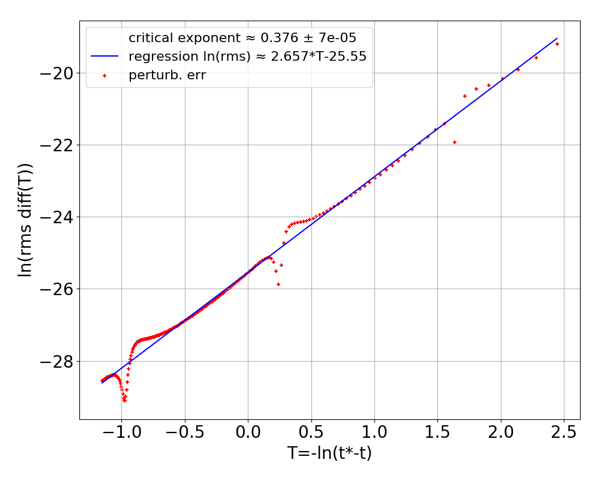}
\end{subfigure}%
\begin{subfigure}{0.34\linewidth}
    \subcaption{\large{Massive Field = 0.001}}
\includegraphicsdpi{\smalldpi}{width=\linewidth}{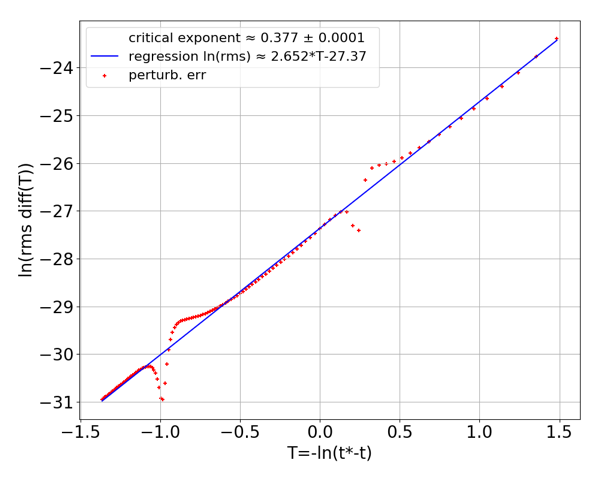}
\end{subfigure}\\
\begin{subfigure}{0.33\linewidth}
    \subcaption{\large{Massive Field = 0.00111}}
\includegraphicsdpi{\smalldpi}{width=\linewidth}{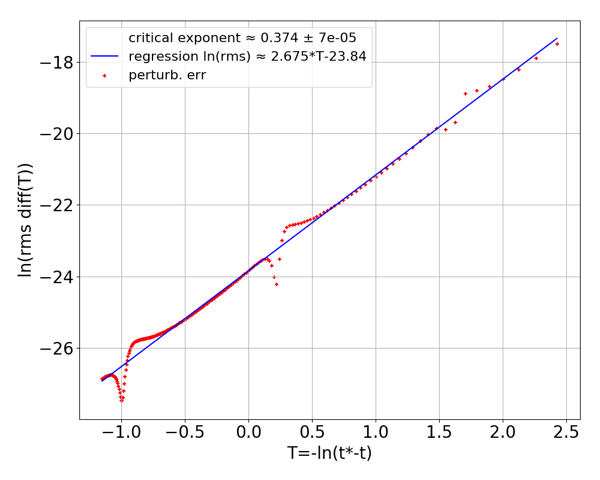}
\end{subfigure}%
\begin{subfigure}{0.33\linewidth}
    \subcaption{\large{Massive Field = 0.001115}}
\includegraphicsdpi{\smalldpi}{width=\linewidth}{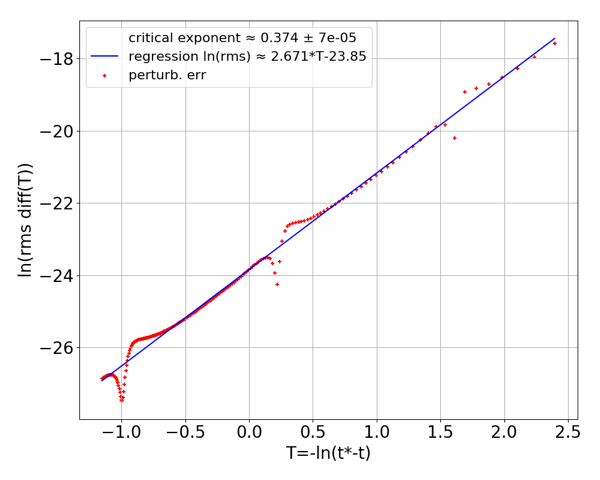}
\end{subfigure}%
\begin{subfigure}{0.34\linewidth}
    \subcaption{\large{Massive Field = 0.0011156}}
\includegraphicsdpi{\smalldpi}{width=\linewidth}{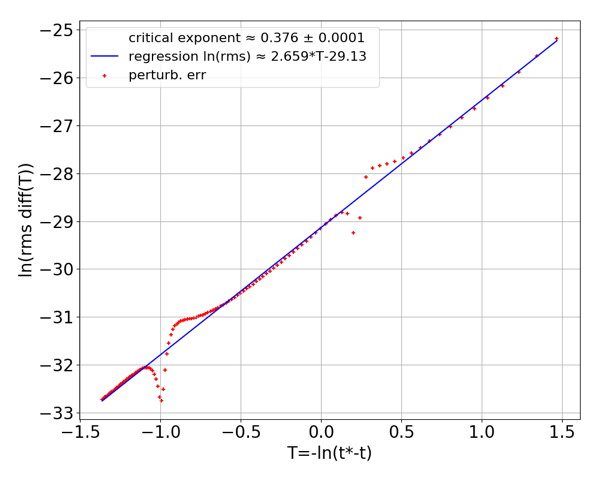}
\end{subfigure}\\
   \caption{\label{fig:typeIIperturbs} Perturbative analyses performed on pairs of near-critical data arising from simulations near the Type~II section of the collapse/dispersal boundary. Each panel represents data taken with the amplitude of the massive field fixed, while the massless field's varies. The values for $\upgamma_{\rom{2}}$ reasonably match their counterparts in Table~\ref{table:regII}, validating our results.}
\end{figure*}

\begin{figure*}[!ht]
\centering
\includegraphicsdpi{\smalldpi}{width=\linewidth}{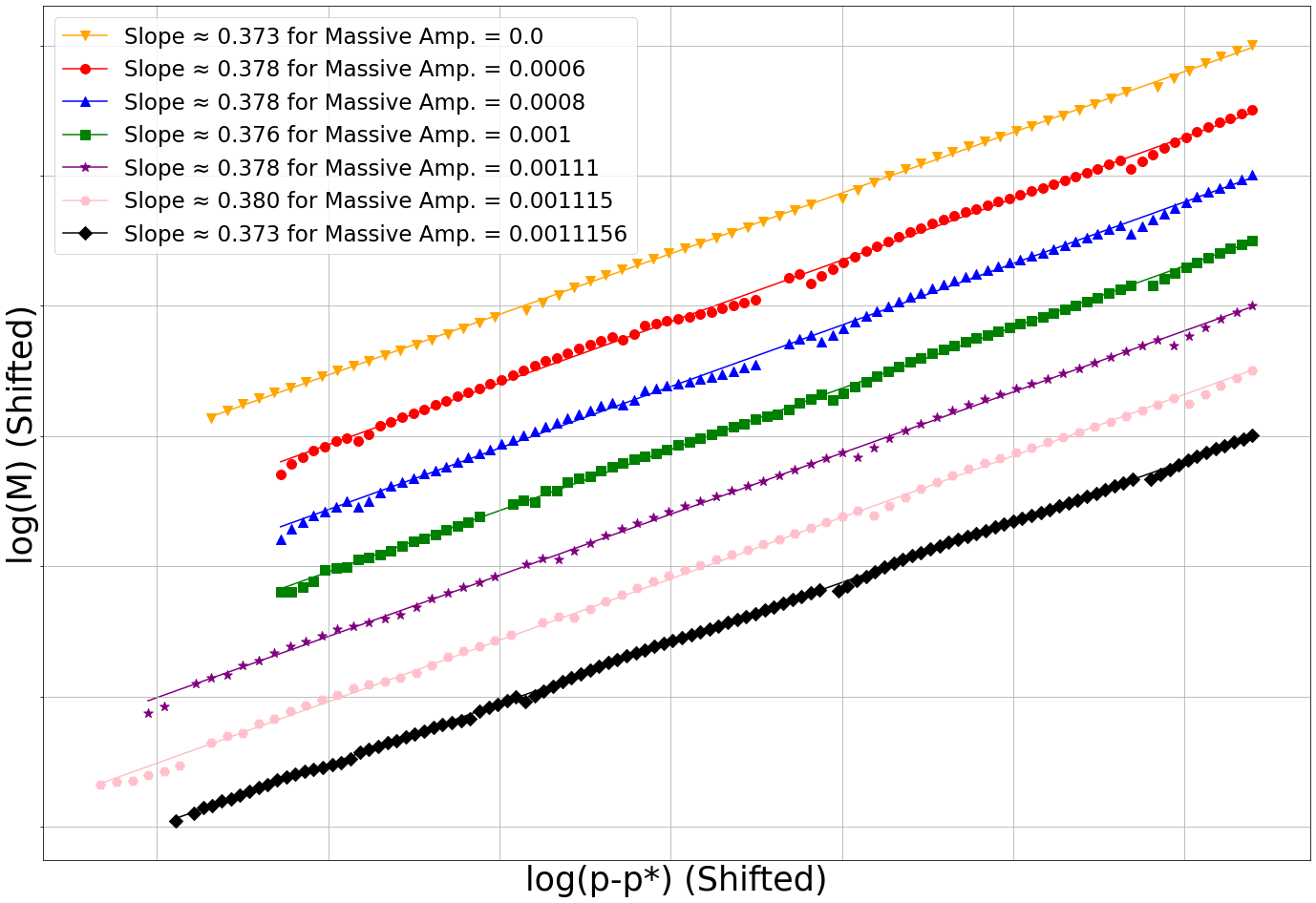}
   \caption{\label{fig:typeIImergereg} A compilation of the linear regression analyses performed on the masses of black holes formed near the Type~II section of the dispersal/collapse boundary. Each set of points cocolored has been shifted uniformly, keeping the slope intact, so as to better present all slopes for comparison by eye. The points corresponding to fixed massive amplitudes 0.00111,0.001115, and 0.0011156 correspond to the red tracts in Fig.~\ref{fig:combinedcolor}.}
\end{figure*}

 Table~\ref{table:regII} and Fig.~\ref{fig:typeIIperturbs} illustrate this point. Each entry in Table~\ref{table:regII} reflects a log-log regression of black hole formation mass versus the massless field amplitude's deviation from criticality, while each panel in Fig.~\ref{fig:typeIIperturbs} shows a perturbative regression analysis via a log-log regression of the $L^2$ difference between two simulations near criticality versus logarithmic time-to-collapse. Figure~\ref{fig:typeIImergereg} features all the mass regressions together for ready comparison by eye.
\par
\begin{figure*}[!ht]
\centering
\begin{subfigure}{\linewidth}
  \includegraphicsdpi{\smalldpi}{width=\linewidth}{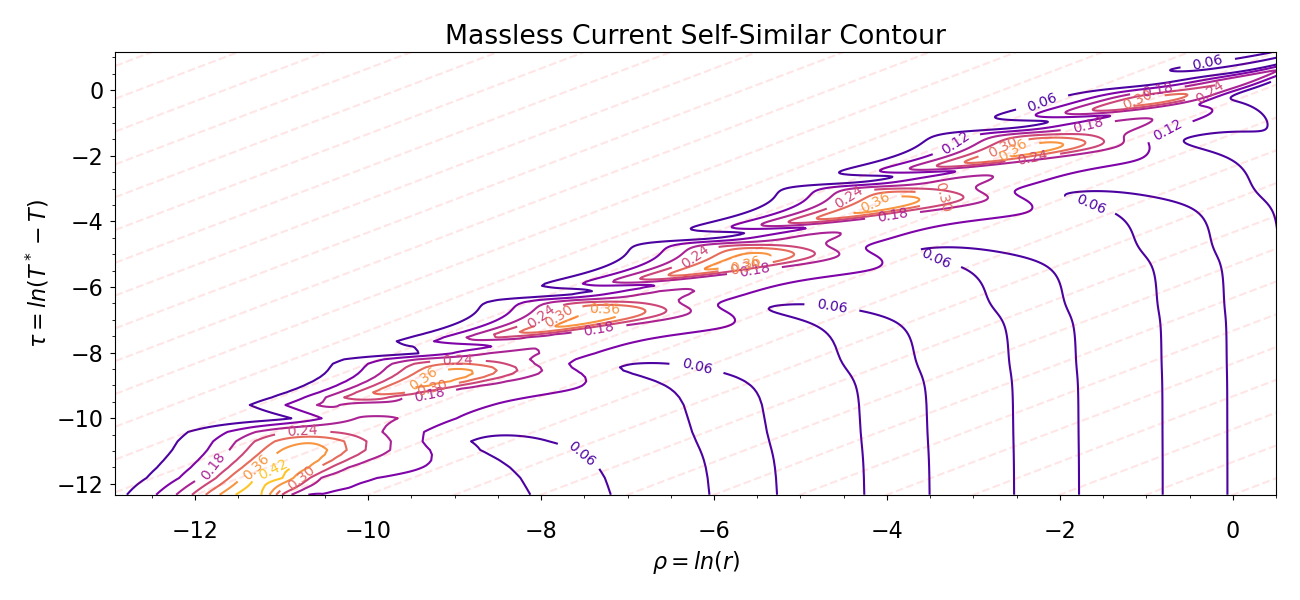}
  \includegraphicsdpi{\smalldpi}{width=\linewidth}{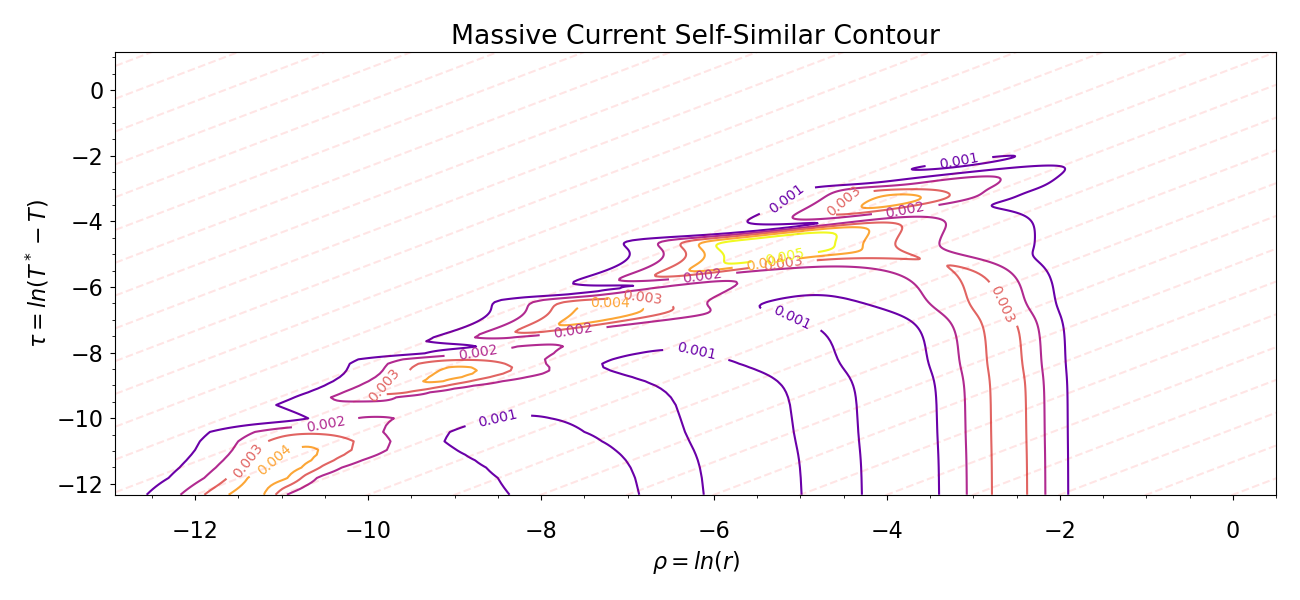}
\end{subfigure}
% \caption{\label{fig:echoing} An example of the expected echoing effect for the massless field across the Type~II portion of the boundary with massive field set to 0.0008. The expected period of $\Delta\approx 3.4$ is observed, as well as the development of similar oscillations in the massive field.}
 \caption{\label{fig:echoing} Contour plots of the partial currents showing the expected echoing effect across the Type~II portion of the boundary. Time here is obtained via integration of the lapse interpolated to the origin of the staggered grid. The massive field's amplitude has been set to 0.0008. Being functions of the fields squared, the partial currents exhibit the expected period of half the conventional $\Delta\approx 3.4$. This can be seen by eye, following a diagonal red dashed line from some neighborhood of $J(\rho,\tau)$ and finding similar features around $J(\rho-\Delta/2,\tau-\Delta/2)$. What is remarkable is the development of weak echoing in the massive current -- this is a consequence of strong dynamical effects controlling the gravitational interaction.}
\end{figure*}

 The same value of the dominant exponent, $\upgamma_{\rom{2}} \approx 2.66$, is observed within reasonable precision across the board. This is the same as the reciprocal of the critical exponent as that associated with the case of the pure massless field from the classic investigation by Choptuik~\cite{choporiginal}. We find the expected echoing effect not only in the massless field, but the emergence of a similar period in the massive field as well. This is depicted in Fig.~\ref{fig:echoing} in terms of the partial currents associated with the Kodama vector.
\par Thus, the presence of the massive field has only a quantitative effect on evolutionary dynamics near criticality. It does not qualitatively alter the fundamental mechanisms near collapse as reflected in the dominant exponent $\upgamma_{\rom{2}}$, although naturally the secondary field continues to fall in after the lapse collapses and contributes to the developing black hole mass.

\begin{table}[ht!]
\centering
\begin{tabular}{ |p{0.24\linewidth}||p{0.15\linewidth}|p{0.15\linewidth}|p{0.26\linewidth}| }
\hline
\multicolumn{4}{|c|}{Type II Collapse Mass Regressions}\\
\hline
Massive Field Amplitude &\multicolumn{1}{c|}{Slope} &\multicolumn{1}{c|}{Intercept} &Dominant\newline Exponent ($\upgamma_{\rom{2}}$)\\
\hline
\multicolumn{1}{|l||}{0.0} &\multicolumn{1}{r|}{0.373} &\multicolumn{1}{r|}{0.248} &\multicolumn{1}{r|}{2.68}\\
\multicolumn{1}{|l||}{0.0006} &\multicolumn{1}{r|}{0.378} &\multicolumn{1}{r|}{0.308} &\multicolumn{1}{r|}{2.64}\\
\multicolumn{1}{|l||}{0.0008} &\multicolumn{1}{r|}{0.378} &\multicolumn{1}{r|}{0.317} &\multicolumn{1}{r|}{2.65}\\
\multicolumn{1}{|l||}{0.001} &\multicolumn{1}{r|}{0.376} &\multicolumn{1}{r|}{0.322} &\multicolumn{1}{r|}{2.66}\\
\multicolumn{1}{|l||}{0.00111} &\multicolumn{1}{r|}{0.378} &\multicolumn{1}{r|}{0.315} &\multicolumn{1}{r|}{2.67}\\
\multicolumn{1}{|l||}{0.001115} &\multicolumn{1}{r|}{0.380} &\multicolumn{1}{r|}{0.315} &\multicolumn{1}{r|}{2.67}\\
\multicolumn{1}{|l||}{0.0011156} &\multicolumn{1}{r|}{0.373} &\multicolumn{1}{r|}{0.269} &\multicolumn{1}{r|}{2.68}\\
\hline
\end{tabular}
   \caption{Linear regression analyses performed on the masses of black holes formed along the Type~II section of the collapse/dispersal boundary. Each row represents data taken with the amplitude of the massive field fixed, while the massless field's amplitude varies. All values for $\upgamma_{\rom{2}}$ agree with that of the pure massless scalar field}
\label{table:regII}
\end{table}

\subsection{Type~I Boundary}
%In contrast to behavior at the Type~II boundary, the presence of a second field has a significant effect upon Type~I collapse. The presence of the massless field leads to the appearance of complex and difficult to numerically tame dynamics away from the triple point and away from the case of the pure massive field. This manifests in the behavior of the dominant mode associated to the Type~I critical solution, which here in fact changes depending upon the massless field amplitude.
In contrast to the Type~II section of the boundary, the presence of an additional field has a significant effect upon Type~I collapse. This manifests quantitatively in the behavior of the dominant exponent associated to the Type~I critical solution, which we have found varies depending on the massless field amplitude.
%\par Figures \ref{fig:typeIregs} and \ref{fig:typeIperturbs} communicate this point similarly to their paired counterparts \textit{supra}. Each panel in Fig.~\ref{fig:typeIregs} depicts a linear regression taken with respect to the time to the collapse of the lapse arising from black hole formation near the Type~I/dispersal boundary. The massless field amplitude is held fixed, while the abscissa reflects the logarithm of the massive amplitude's deviation from criticality. Again, these regressions are compiled in Fig~\ref{fig:typeImergereg} for ready comparison by eye. The panels collected in Fig.~\ref{fig:typeIperturbs} show various perturbative analyses with data taken near criticality for the massive field, with the massless field amplitude fixed to the value equaling its partner panel in Fig.~\ref{fig:typeIregs}. 
\par
\begin{figure*}[!ht]
\centering
\begin{subfigure}{0.33\linewidth}
    \subcaption{\normalsize{Massless Field = 0.0 (I)}}
\includegraphicsdpi{\smalldpi}{width=\linewidth}{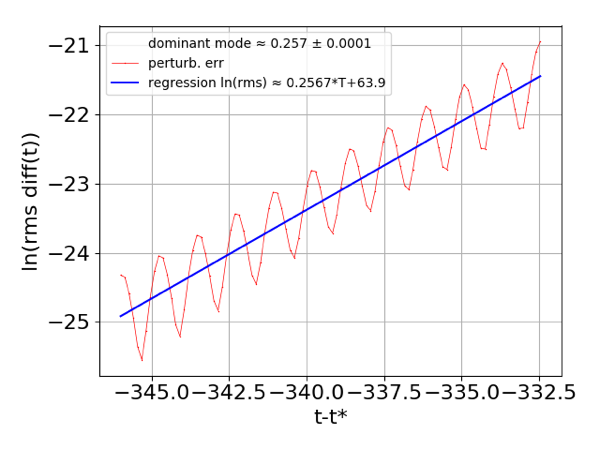}
\end{subfigure}%
\begin{subfigure}{0.33\linewidth}
    \subcaption{\normalsize{Massless  Field = 0.0 (II)}}
\includegraphicsdpi{\smalldpi}{width=\linewidth}{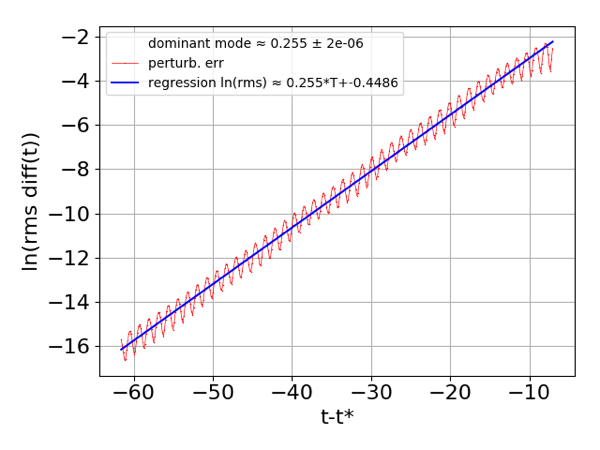}
\end{subfigure}%
\begin{subfigure}{0.33\linewidth}
    \subcaption{\normalsize{Massless Field = 0.04 (I)}}
\includegraphicsdpi{\smalldpi}{width=\linewidth}{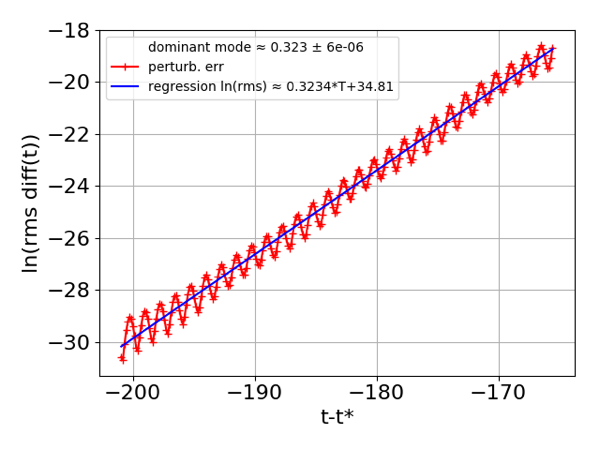}
\end{subfigure}\\
\begin{subfigure}{0.33\linewidth}
    \subcaption{\normalsize{Massless  Field = 0.04 (II)}}
\includegraphicsdpi{\smalldpi}{width=\linewidth}{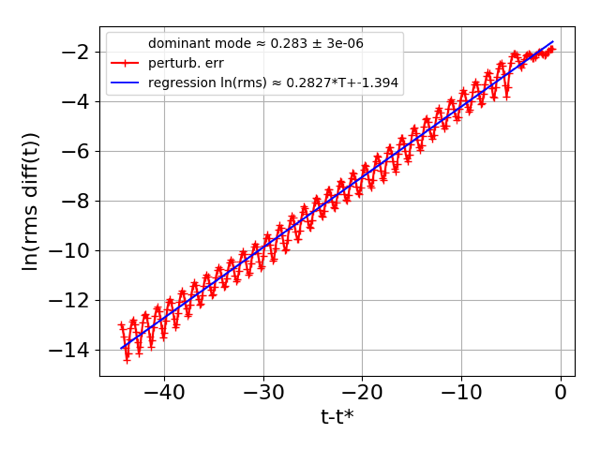}
\end{subfigure}%
\begin{subfigure}{0.33\linewidth}
    \subcaption{\normalsize{Massless Field = 0.043}}
\includegraphicsdpi{\smalldpi}{width=\linewidth}{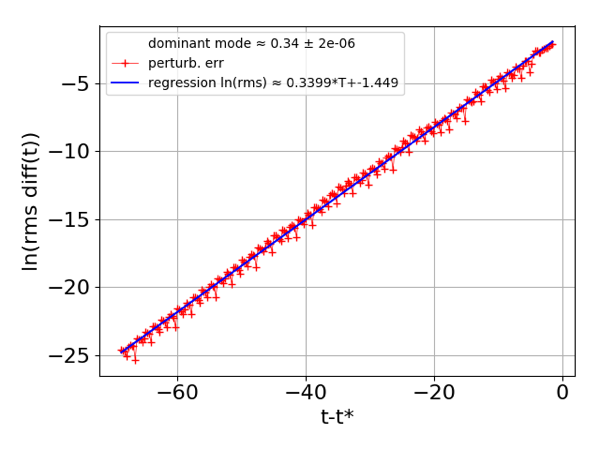}
\end{subfigure}%
\begin{subfigure}{0.33\linewidth}
    \subcaption{\normalsize{Massless Field = 0.0434}}
\includegraphicsdpi{\smalldpi}{width=\linewidth}{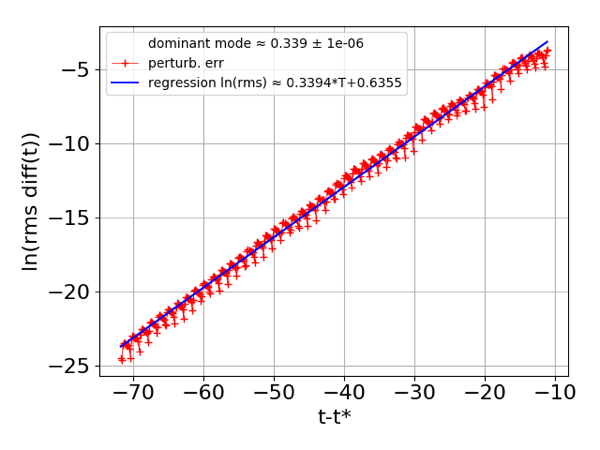}
\end{subfigure}
%   \caption{\label{fig:typeIperturbs} Perturbative analyses performed on pairs of near-critical data arising from simulations near the Type~I/dispersal boundary. Each panel represents data taken with the amplitude of the massless field fixed, while the massive field's varies. Each plot corresponding in position with its collapse of the lapse regression counterpart in Fig~\ref{fig:typeIregs} \textit{supra}. The values for $\gamma$ reasonably match their counterparts, validating our dynamical analysis.}
   \caption{\label{fig:typeIperturbs} Perturbative analyses performed on pairs of near-critical data arising from simulations near the Type~I section of the collapse/dispersal boundary. Each panel represents data taken with the amplitude of the massless field fixed, while the massive field's varies. The values for $\upgamma_{\rom{1}}$ reasonably match their counterparts in Table~\ref{table:regI}, validating our results.}
\end{figure*}

\begin{figure*}[!ht]
\centering
\includegraphicsdpi{\smalldpi}{width=\linewidth}{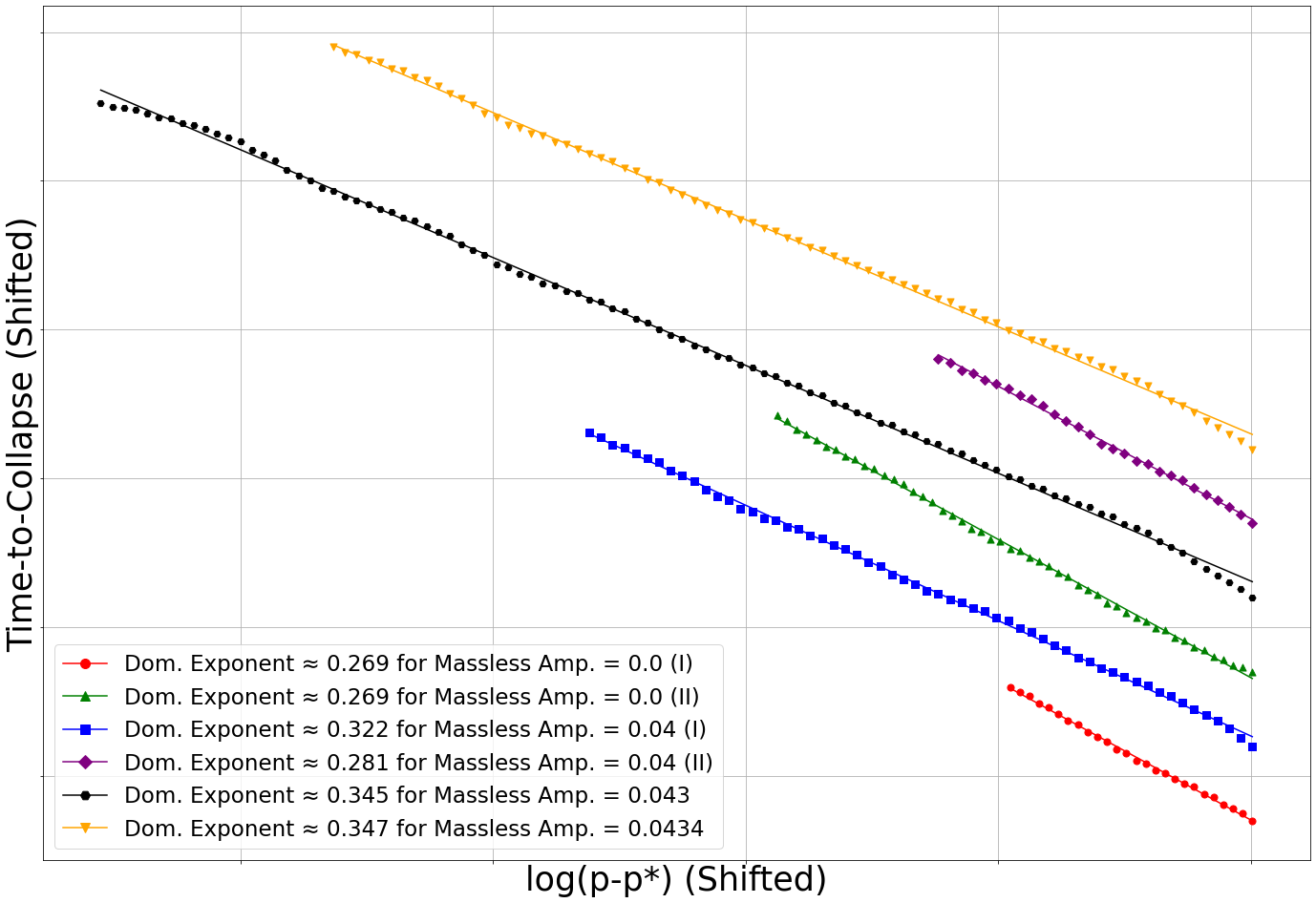}
   \caption{\label{fig:typeImergereg} A compilation of the linear regression analyses performed on the asymptotic times-to-collapse of the lapse along the Type~I section of the dispersal/collapse boundary. Sets of points of the same color have been shifted uniformly, keeping their regression intact, so as to better present all slopes for comparison by eye. The variation in slope is readily observed. The points with fixed massless amplitudes 0.04 and 0.043 correspond to the yellow tracts in Fig.~\ref{fig:combinedcolor}.}
\end{figure*}

 Table~\ref{table:regI} and Fig.~\ref{fig:typeIperturbs} communicate this point similarly to their foregoing Type~II counterparts. The rows of Table~\ref{table:regI} give the results of log-log regressions of the asymptotic time-to-collapse versus the massive field amplitude's deviation from criticality, while the panels collected in Fig.~\ref{fig:typeIperturbs} show various perturbative analyses performed via log-log regressions of the $L^2$ difference between two simulations near criticality versus time-to-collapse. Fig.~\ref{fig:typeImergereg} compiles all the time-to-collapse regressions to better facilitate comparison by eye: in contrast to Fig.~\ref{fig:typeIImergereg}, there appear to be two distinct slope values.
\par
\begin{figure}[!ht]
\centering
\begin{subfigure}{\linewidth}
\includegraphicsdpi{\smalldpi}{width=\linewidth}{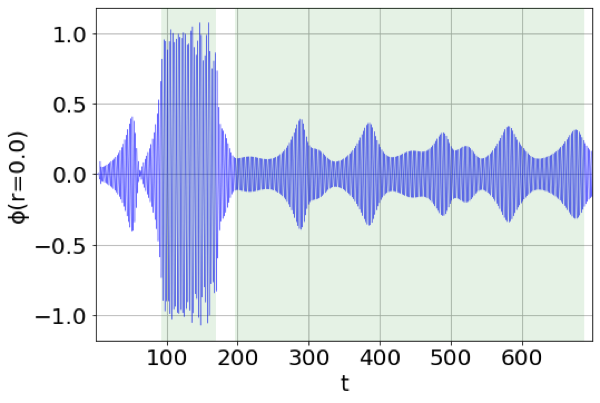} \\
\includegraphicsdpi{\smalldpi}{width=\linewidth}{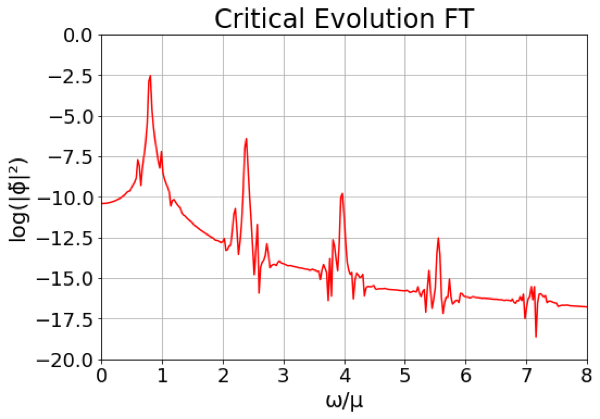}\\
\includegraphicsdpi{\smalldpi}{width=\linewidth}{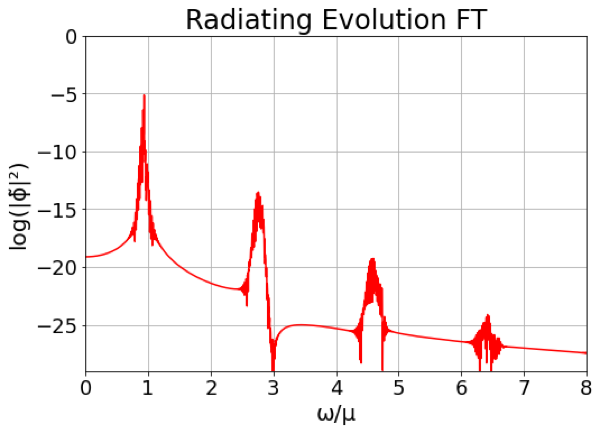}
\end{subfigure}
 \caption{\label{fig:period} The behavior of the massive field versus time elapsed at the origin and Fourier decompositions thereof during different regimes. The scenario is slightly subcritical. The two shaded regions in the top figure are the two regions subjected to Fourier analysis shown in the lower two. Critical evolution features peaks at $\approx 0.80\mu^{-1}$ and odd harmonics with sidebands, while dispersal behavior exhibits sharper peaks at $\approx 0.94\mu^{-1}$ with weaker harmonics.}

%The two left figures correspond to critical evolution, while the right two correspond to late-time dispersal.
\end{figure}

 We observe the expected periodicity of the underlying critical solution across the Type~I portion of the boundary. Taking a representative case in Fig.~\ref{fig:period}, we find that, during critical evolution, the field oscillates with frequency $\approx 0.80\mu^{-1}$, with excited odd harmonics and sidebands corresponding to the slower undulations. Meanwhile, during the ``radiating'' period, the frequency changes to $\approx 0.94\mu^{-1}\approx \mu^{-1}$, with comparatively less excited harmonics, all in fair agreement with previous studies~\cite{othermassive,suen,suen2}.
\par In contrast to the Type~II case, the value of the dominant exponent $\upgamma_{\rom{1}}$ noticeably changes. Near the triple point, it takes on the value $\upgamma_{\rom{1}} \approx 0.34$. However, near the regime of a pure massive field, $\upgamma_{\rom{1}}$ takes on a value closer to $\approx 0.27$. Moreover, we note a significant decrease in the typical time-to-collapse as we near the triple point. This second fact matches the intuition that frustrated multicriticality otherwise runs against: collapse still happens ``earlier,'' as we might expect due to the presence of extra matter in the form of the secondary field, despite the counterintuitive inhibition of collapse seen in the shift in criticality.
%\par The first panel of Fig~\ref{fig:typeIregs} is in fact a zoom-in of a larger graph of times-to-collapse in the case of the pure massive field, which is shown in its entirety in the first panel of Fig~\ref{fig:jumps}. Generally we observe, even moderately away from the triple point, apparent jumps in the time-to-collapse in the Type~I case. Whether these jumps occur near the triple point at greater precision has not been determined. 
\par
\begin{figure*}[!ht]
\centering
\includegraphicsdpi{\smalldpi}{width=0.45\linewidth}{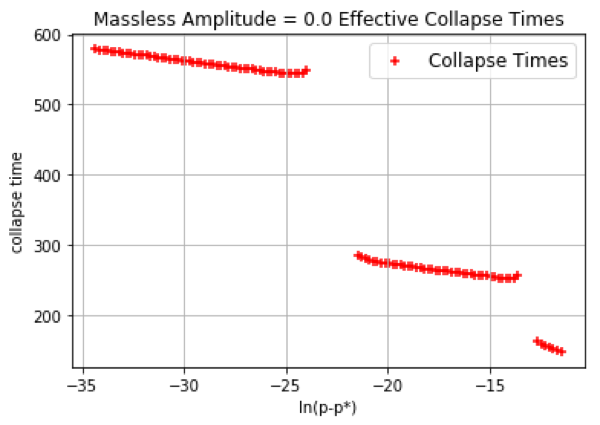}
\includegraphicsdpi{\smalldpi}{width=0.45\linewidth}{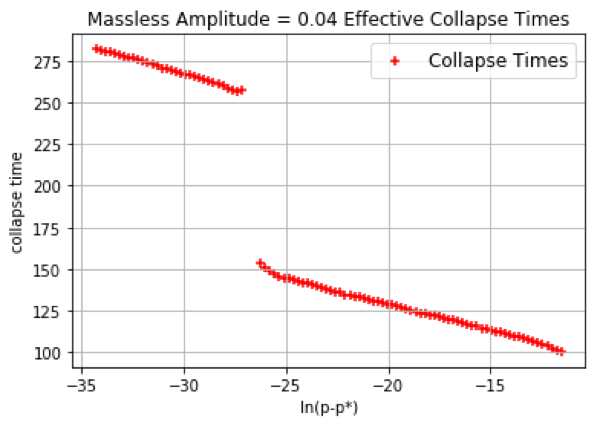}
   \caption{\label{fig:jumps} Across the Type~I section of the dispersal/collapse boundary, jumps appear the time-to-collapse of the lapse. The slopes of the two most dominant ridges of the left panel are appreciably the same, while for the right they appreciably differ. This difference is repeated in substance by their respective perturbative analyses in Figs~\ref{fig:zeroshiftingperturbs} and Fig.~\ref{fig:4shiftingperturbs}. While the existence of the jumps is likely merely a quirk of the initial data, whether the slope changes or not is a deeper dynamical feature.}
\end{figure*}

 Some entries and sets in in Table~\ref{table:regI} and Fig.~\ref{fig:typeIperturbs} are labelled by additional suffixes. This is because, as shown in Fig.~\ref{fig:jumps}, distinct and numerically consistent ridges appear in some collapse-of-the-lapse regressions. Even moderately away from the triple point, apparent jumps in the time-to-collapse are observed along the Type~I section of the boundary. The existence of these jumps, we suspect, may be attributed to a quirk of the Gaussian initial data. Whether these jumps completely disappear or not near the triple point has not been determined. 

\begin{table}[ht!]
\centering
\begin{tabular}{ |p{0.24\linewidth}||p{0.15\linewidth}|p{0.15\linewidth}|p{0.26\linewidth}| }
\hline
\multicolumn{4}{|c|}{Type I Collapse Time Regressions}\\
\hline
%Massless Field Amplitude &Slope &Intercept &Dominant Mode ($\gamma$)\\
Massless Field Amplitude &\multicolumn{1}{c|}{Slope} &\multicolumn{1}{c|}{Intercept} &Dominant\newline Exponent ($\upgamma_{\rom{1}}$)\\
\hline
\multicolumn{1}{|l||}{0.0 (I)} &\multicolumn{1}{r|}{-3.72} &\multicolumn{1}{r|}{198.} &\multicolumn{1}{r|}{0.269}\\
\multicolumn{1}{|l||}{0.0 (II)} &\multicolumn{1}{r|}{-3.72} &\multicolumn{1}{r|}{451.} &\multicolumn{1}{r|}{0.269}\\
\multicolumn{1}{|l||}{0.04 (I)} &\multicolumn{1}{r|}{-3.10} &\multicolumn{1}{r|}{66.9} &\multicolumn{1}{r|}{0.322}\\
\multicolumn{1}{|l||}{0.04 (II)} &\multicolumn{1}{r|}{-3.56} &\multicolumn{1}{r|}{160.} &\multicolumn{1}{r|}{0.281}\\
\multicolumn{1}{|l||}{0.043} &\multicolumn{1}{r|}{-2.90} &\multicolumn{1}{r|}{66.8} &\multicolumn{1}{r|}{0.345}\\
\multicolumn{1}{|l||}{0.0434} &\multicolumn{1}{r|}{-2.88} &\multicolumn{1}{r|}{65.8} &\multicolumn{1}{r|}{0.347}\\
%0.0 (I) &-3.72 &198. &0.269\\
%0.0 (II) &-3.72 &451. & 0.269\\
%0.04 (I) &-3.10 &66.9 &0.322\\
%0.04 (II) &-3.56 &160. &0.281\\
%0.043 &-2.90 &66.8 & 0.345\\
%0.0434 &-2.88 &65.8 & 0.347\\
\hline
\end{tabular}
   \caption{Linear regressions performed on times-to-collapse of the lapse for black holes forming along the Type~I section of the collapse/dispersal boundary. Each entry represents data taken with the amplitude of the massless field fixed, while the massive field's varies. $\upgamma_{\rom{1}}$ appears to take on a different value depending on whether the massless field amplitude places the scenario near triple point, or whether it's near the pure massive field case, and may vary across different ridges of linearity appearing in the data.}
\label{table:regI}
\end{table}

%\begin{figure*}[!ht]
%\centering
%\includegraphics[width=0.3\linewidth]{0.png}
%\includegraphics[width=0.3\linewidth]{43.png}
%\includegraphics[width=0.3\linewidth]{434.png}
%   \caption{\label{fig:typeIregs} Linear regressions performed on time-to-collapse of the lapse arising from black holes forming along the Type~I/dispersal boundary. Each panel represents data taken with the amplitude of the massless field fixed, while the massive field's varies. $\gamma$ appears to take on a different value depending on whether the massless field amplitude places the scenario near triple point, or whether it's near the pure massive field case.}
%\end{figure*}

%\begin{figure*}[!ht]
%\centering
%\includegraphics[width=0.45\linewidth]{zerofieldjumps.png}
%\includegraphics[width=0.45\linewidth]{4total.png}
%   \caption{\label{fig:jumps} Across the Type~I section of the dispersal/collapse boundary, jumps appear the time-to-collapse of the lapse. The slopes of the two most dominant ridges of the left panel are appreciably the same, while for the right they appreciably differ. This difference is repeated in substance by their respective perturbative analyses in Figs~\ref{fig:zeroshiftingperturbs} and Fig.~\ref{fig:4shiftingperturbs}.}
%\end{figure*}

\par
\begin{figure*}[!ht]
\centering
\begin{subfigure}{\linewidth}
\includegraphicsdpi{\smalldpi}{width=0.45\linewidth}{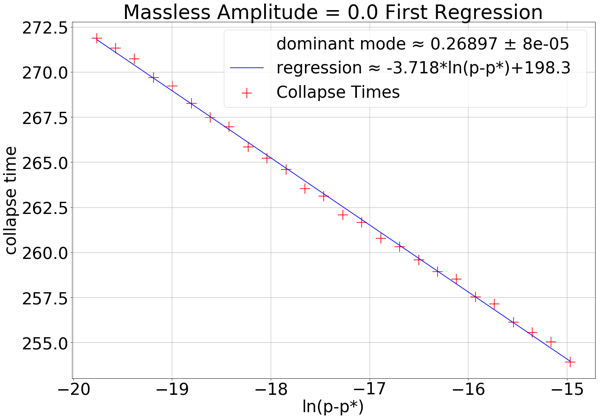}
\includegraphicsdpi{\smalldpi}{width=0.45\linewidth}{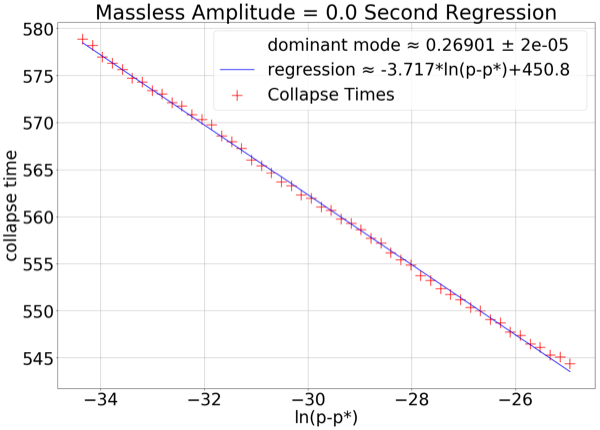}
\end{subfigure}
   \caption{\label{fig:4jumpzoom} Zoom-in regressions on the individual ridges seen in the left panel of Fig.~\ref{fig:jumps}. The equal dominant exponents derived from these individual ridges coincide with the exponents derived from their respective intervals of perturbative evolution in Fig.~\ref{fig:zeroshiftingperturbs} \textit{supra}.}
\end{figure*}

 While the two most prominent slopes in the left panel of Fig.~\ref{fig:jumps} appear to be equal, the two slopes in the panel on the right slightly differ. This variance appears to not be numerical error, as it is reflected in the perturbative analysis in Fig.~\ref{fig:4shiftingperturbs}. The top plot of Fig.~\ref{fig:4shiftingperturbs} depicts a sort of rotation of the right panel of Fig.~\ref{fig:jumps}. The evolution of the perturbation passes through distinct regimes, with different dominant slopes modulated by underlying undulations. The two intervals of greatest growth are depicted in the lower two plots of Fig.~\ref{fig:4shiftingperturbs}. The dominant exponents obtained therefrom appreciably match those derived from the individual ridges of the right panel of Fig.~\ref{fig:jumps}, which are depicted in detail in Fig.~\ref{fig:4jumpzoom}, covering ordinate intervals roughly commensurate with the abscissa intervals of their counterparts. The two apparent dominant exponents have approximate values $0.32$ and $0.28$
\par
\begin{figure*}[!hb]
\centering
\begin{subfigure}{\linewidth}
\includegraphicsdpi{\smalldpi}{width=0.45\linewidth}{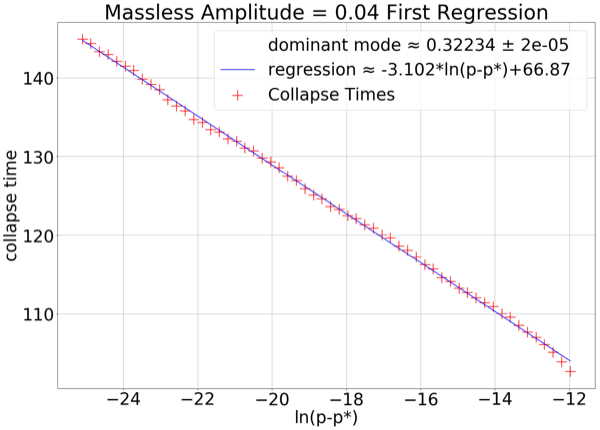}
\includegraphicsdpi{\smalldpi}{width=0.45\linewidth}{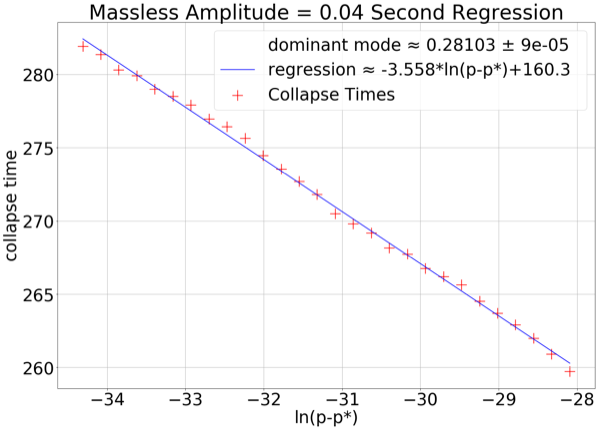}
\end{subfigure}
  \caption{\label{fig:zerojumpzoom} Zoom-in regressions on the individual ridges seen in the right panel of Fig.~\ref{fig:jumps}. The dominant exponents derived from these individual ridges coincide with the exponents derived from their respective intervals of perturbative evolution in Fig.~\ref{fig:4shiftingperturbs} \textit{infra}.}
\end{figure*}

\begin{figure}[htbp]
  \centering
    \includegraphicsdpi{\smalldpi}{width=0.9\linewidth}{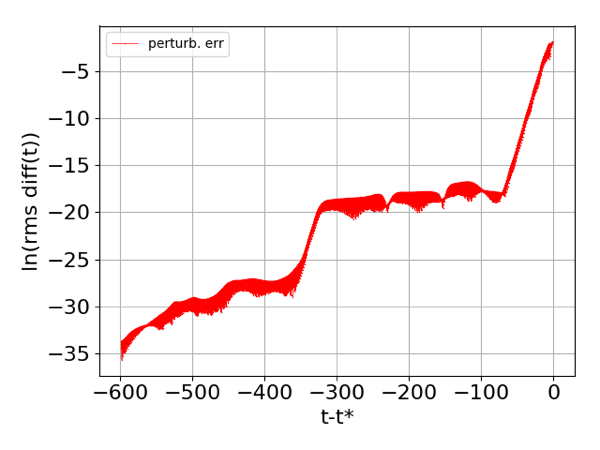}\\
\includegraphicsdpi{\smalldpi}{width=0.8\linewidth}{zeroIregrrpi64000.png}\\
\includegraphicsdpi{\smalldpi}{width=0.8\linewidth}{zeroIIregrrpi64000.png}
 \caption{\label{fig:zeroshiftingperturbs} Evolution of a perturbation away from the critical solution along the Type~I section of the collapse/dispersal boundary in the case of the pure massive field. The top plot depicts the time evolution over the course of the entire simulation, and illustrates how the perturbation undergoes different evolutionary regimes. The lower plots zoom in on the two intervals of greatest growth. Their appreciably equal slopes correspond to equal dominant exponents.}
\end{figure}

\begin{figure}[htbp]
\includegraphicsdpi{\smalldpi}{width=0.9\linewidth}{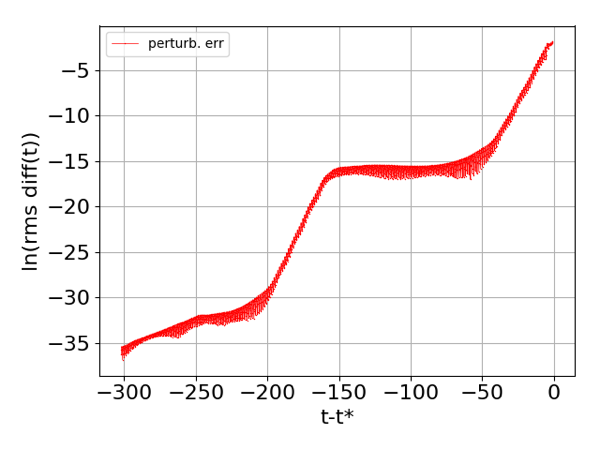}\\
\includegraphicsdpi{\smalldpi}{width=0.8\linewidth}{4Iregrrpi64000.png}\\
\includegraphicsdpi{\smalldpi}{width=0.8\linewidth}{4IIregrrpi64000.png}
    \caption{\label{fig:4shiftingperturbs} Evolution of a perturbation away from the critical solution along the Type~I section of the collapse/dispersal boundary with fixed massless field amplitude = 0.04. The top plot depicts the time evolution over the course of the entire simulation, and illustrates how the perturbation exhibits different regimes. The bottom two plots zoom in on the two intervals of greatest growth, whose differing slopes show that the dominant perturbation undergoes a subtle changes.}
\end{figure}

 The equivalent contrasting story expanding upon the left panel of Fig.~\ref{fig:jumps} is present in Figs~\ref{fig:zerojumpzoom} and Fig.~\ref{fig:zeroshiftingperturbs}. In this case, as previously said, the slopes are approximately equal, tending around $0.26$.

\section{Transfer of Energy}
\par In the introduction it was mentioned that the phenomenology we observe might be understood intuitively in terms of a transfer of energy. While this approach may not be helpful for its lack of decidability, in that a simple hypothesis can just as easily guess at a catalyzation of collapse instead of the inhibition we observe, it is nevertheless an interesting question to ask what this behavior actually is.
\par Although most local definitions of energy in general relativity suffer from various deficiencies in the absence of any particular symmetry, in the simple spherical there exists a conserved flux given as the contraction of the stress-energy tensor with a vector field known as the Kodama vector~\cite{kodama}. This is a kind of replacement for the fluxes associated with Killing-vectors. For the general spherically-symmetric line element $ds^2=g_{ab}dx^adx^b+R^2d\Omega^2$ its nonzero components may be written as $K^a=\epsilon^{ab}\partial_bR$, where $ab$ are indices for the two-dimensional metric $g_{ab}$. Here it takes the simple form $\vec{K}=\tfrac{1}{a\alpha}\vec{\partial_t}$. Contracting with the stress-energy tensor provides the conserved current
\begin{align}
  J^\mu &= {T^\mu}_\nu K^\nu, \nonumber \\
  &= \tfrac{1}{a\alpha}{T^\mu}_0, \nonumber \\
  &\propto \tfrac{1}{a\alpha}\sum\limits_i\Big[\partial^\mu\phi_i\partial_0\phi_i-\tfrac{1}{2a^2}{\delta^\mu}_0(\pi_i^2-\psi_i^2-m_i^2a^2\phi_i^2)\Big]\nonumber
\end{align}
which yields, for the current case of interest,
\begin{align}
  J^0\propto\tfrac{1}{2a^3\alpha}\sum\limits_i\left(\pi_i^2+\psi_i^2+m_i^2a^2\phi_i^2 \right). \numberthis
\end{align}
\vfill\null
\par We can now integrate this over a spacelike leaf of the ADM foliation to obtain the conserved charge
\begin{align*} \label{eq:charges}
  Q_J&=\int_V(*J), \nonumber \\
  &\propto\int\tfrac{\alpha}{2a^3}\sum\limits_i\left(\pi_i^2+\psi_i^2+m_i^2a^2\phi_i^2 \right)\tfrac{\sqrt{|g|}}{\alpha^2}drd\theta d\varphi, \nonumber \\
  &\propto\int_V\sum\limits_i\tfrac{r^2}{a^2}\left(\pi_i^2+\psi_i^2+m_i^2a^2\phi_i^2 \right)dr. \numberthis
\end{align*}
\par The above expression, after applying the equations of motion, leads to the classic Misner-Sharp mass function $M=\tfrac{r}{2}(1-a^{-2})$~\cite{misnersharp}. The thing of value obtained here is how this derivation provides a natural splitting of the mass aspect into a sum of the various fields' contributions to the stress-energy tensor. While these terms individually are not conserved, they nevertheless sum to a conserved charge, and so they can be used to monitor how energy moves between matter sources. Even without this machinery, this idea is well-known, and has been used as a diagnostic before by other authors~\cite{chopboson}.
\par
\begin{figure*}[!ht]
\centering
\begin{subfigure}{\linewidth}
\includegraphicsdpi{\smalldpi}{width=0.48\linewidth}{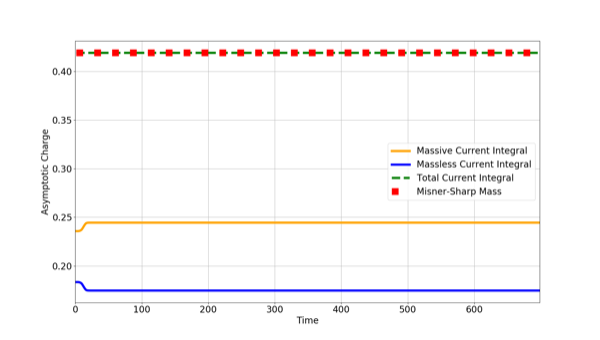}
\includegraphicsdpi{\smalldpi}{width=0.48\linewidth}{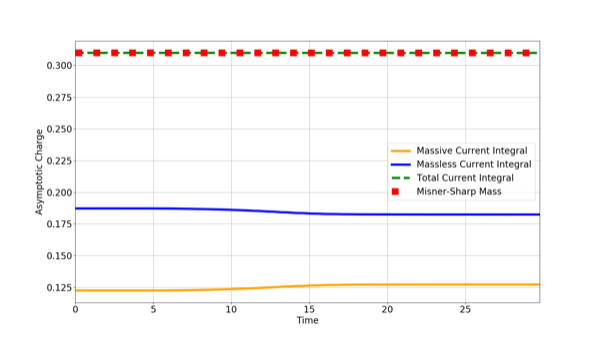}
\end{subfigure}
  \caption{\label{fig:chargeplottotal}Graphs of the total and constituent charges and Misner-Sharp mass corresponding to two slightly subcritical scenarios across the Type~I/- and Type~II/dispersal parts of the phase diagramFig.~\ref{fig:combinedcolor}. The left is subcritical with the massless amplitude equal 0.043, while the right is subcritical with the massive amplitude equal to 0.0008}
\end{figure*}

 As an illustrative example, we show in Fig.~\ref{fig:chargeplottotal} graphs of the various currents for two scenarios across the Type~I and Type~II portions of the boundary in Fig.~\ref{fig:combinedcolor}. We numerically observe, as claimed \textit{supra}, that the two partial charges corresponding to integrals of the individual summands of Eq.~(\ref{eq:charges}) sum to the conserved total Misner-Sharp mass aspect. We not only obtain a quantitative value that can be put to the transfer of energy mentioned before, but also observe a distinct curiosity in how, for both scenarios, the massive field charge increases in time at the expense of the massless field.
\par This is similar to phenomenon observed by Hawley and Choptuik~\cite{chopboson}, but where there the authors considered a small massless field perturbing a complex massive boson star, here both fields are present in significant extent, the massless charge in fact being greater than the massive charge in the right graph of the figure. Despite this, a sizable portion of the energy is still nevertheless transferred around the time of the Type~II critical evolution to the massive field.
\par More interesting is how applying a naive intuition to consideration of these currents falls short. Comparing the subcritical mixed field Type~I scenario in the left panel of Fig.~\ref{fig:chargeplottotal} with the supercritical case in Fig.~\ref{fig:chargesolemassive} shows that scenarios with not only greater total charge, but even greater partial charges may not necessarily have a greater propensity to collapse.
%Guiding one's intuition with this notion of energy transfer, while certainly valid, may fail to yield the correct conclusion, as comparison of the top of Fig.~\ref{fig:chargeplottotal} with Fig.~\ref{fig:chargesolemassive} shows.
%\par The moral that the total charge doesn't tell the whole story. Now, of course it may be the case that a configuration with greater total charge doesn't collapse while one with less does if the former is more compact. This is not so much the case here, however, as our choice of initial data shows; moreover, looking at the partial currents as a function of position and time, as seen in Fig.~\ref{fig:currents}, reveals rather interesting behavior. The strong gravitational effects occurring during the critical evolution are seen to result in high frequency modes of the massless field transferring to the massive field, which spill outwards and generate daughter modes as they encounter the original massive current. This is a more dramatic manifestation of the transfer of echoing seen in Fig.~\ref{fig:echoing}, and is the likely true culprit of the decrease in compactification that inhibits collapse.
%
\par The idea that a less massive configuration may lead to black hole formation while a more massive one doesn't is hardly inconceivable, since intuitively the relevant notion is density. However, since the initial data for the massive field does not differ significantly between the left panel of Figs.~\ref{fig:chargeplottotal} and \ref{fig:chargesolemassive}, this kind of difference in aspect is inadequate to explain this. A qualitative change in the time evolution due to gravitational coupling with the massless field appears necessary; this is intriguing, since it might be thought that the effect of gravity should be to concentrate the fields together and hence promote, rather than inhibit, collapse.
\par
\begin{figure*}[!ht]
\centering
\begin{subfigure}{\linewidth}
\includegraphicsdpi{\smalldpi}{width=0.48\linewidth}{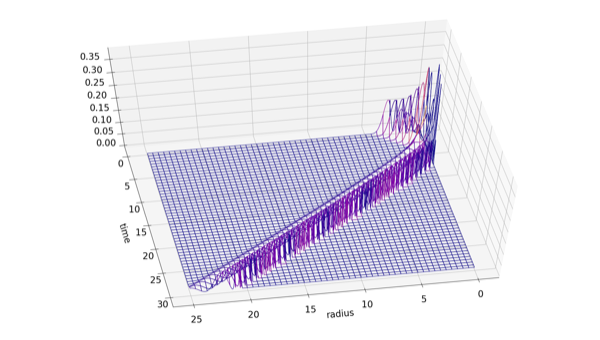}
\includegraphicsdpi{\smalldpi}{width=0.48\linewidth}{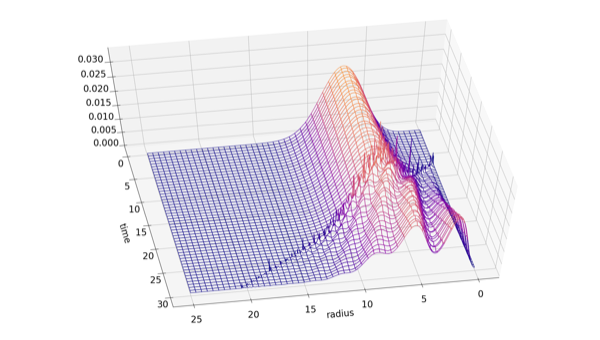}
\end{subfigure}
   \caption{\label{fig:currents}Graphs of the individually non-conserved constituent currents going in Eq.~(\ref{eq:charges}) for a subcritical scenario across the Type~II portion of the boundary. We observe the spontaneous emergence of a high-frequency outgoing mode during the Type~II critical evolution period that, due to the strong coupling occurring near criticality, transfers to the massive current. This is the source of the change in the partial charges.}
\end{figure*}

\begin{figure}[!ht]
\includegraphicsdpi{\smalldpi}{width=\linewidth}{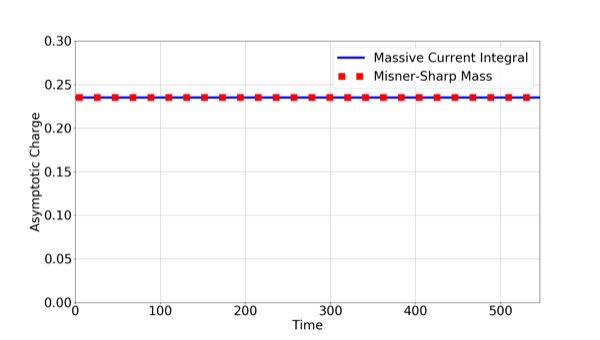}
   \caption{\label{fig:chargesolemassive} Graph of the total current and Misner-Sharp mass for the scenario of a pure slightly supercritical massive field. Comparing with the left of Fig.~\ref{fig:chargeplottotal}, it is seen that although the former has greater total partial charge in the massive field -- and greater total conserved current -- the former nevertheless does not collapse.}
\end{figure}

 A look at the partial currents as a function of position and time in Fig.~\ref{fig:currents} shows behavior that might provide such a mechanism. The strong gravitational effects occurring during critical evolution are seen to result in a transfer of high frequency modes from the massless field to the massive field. These modes proceed to spill outwards and generate daughter modes as they scatter off of the original massive current. This is a more dramatic manifestation of the transfer of echoing seen in Fig.~\ref{fig:echoing}, and is likely the true source of the decrease in compactification, hence the ultimate cause of inhibited collapse.

\section{Discussion}
\par
\begin{figure}[htbp]
  \centering
\includegraphicsdpi{\smalldpi}{width=0.95\linewidth}{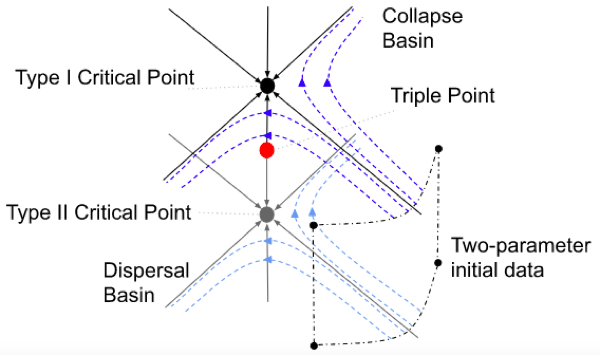}
 \caption{\label{fig:dynobad} A simplified version of the possible dynamical system investigated in this paper. The arrows suggest the local tendency of time evolution through the phase space following the locally dominant perturbation. In addition to the two most prominent influences on the system -- the Type~II Choptuion critical point and a particular metastable solitons star corresponding to the Type~I critical solution -- we detect an additional effect modifying the Type~I critical exponent near the triple point.}
  %Between the interactions of the two big black dots acting in representative capacity for the most obvious parties to the system considered -- the Choptuon Type~II critical point and the family of metastable soliton stars corresponding to the Type~I critical solution -- an additional effect modifying the Type~I critical exponent near the triple point can be detected.}
\end{figure}

 In our previous paper, we postulated that the multicritical phenomena seen in our scenario could reflect an alternative dynamical situation considered by Gundlach et al.~\cite{panic}, which they ultimately dismissed in the case they considered. This possibility is illustrated in Fig.~\ref{fig:dynobad} here, which shares a kinship with Fig.~13 of their paper. Between the two primary attractors at play (the Choptuon and the family of metastable soliton stars) there exists a third critical solution influencing the dynamics with its own dominant perturbative exponent. Our results \textit{supra}, featuring an apparently changing dominant exponent along the Type~I section of the boundary, support this conclusion: the ``competition'' we observe in our setup would appear to feature, as it were, a third belligerent. 
\par Another comparison with Gundlach et al.'s paper is apposite to our point. In Fig.~8 of their paper, they observe ``breaks'' in the apparent critical exponent for their scenario in the region of their phase space where the Yang-Mills field mostly -- but not overwhelmingly -- dominates the scalar field. They ascribe these breaks to a straight-forward change between their two critical solutions, moving from the massless critical solution ($\upgamma_{\rom{2}} \approx 0.37$) to the Yang-Mills critical solution ($\upgamma_{\rom{2}} \approx 0.2$). It might be wondered if the ``jumps'' we observe are similar in nature.
\par We do not construe the evidence such that our results can be attributed so. The dominant exponent along the Type~II section of the collapse/dispersal boundary is $\approx 2.7$, derived from a logarithmic time scale. Meanwhile, the two differing dominant exponents observed along the Type~I section of the boundary appear to range around $\approx 0.32-0.34$ and $\approx 0.26-0.28$, which are derived from a linear time scale operant upon values of asymptotic $t$ orders of magnitude greater than those reached in the Type~II case. By themselves, these differing values of $\upgamma{\rom{1}/\rom{2}}$ aren't directly comparable precisely because they come from analysis based upon different symmetry assumptions; however, considering the Type~I section of the boundary in logarithmic units doesn't yield dominant exponents consistent with the collapse time regressions, or even anything appreciably linear. We accordingly cannot ascribe the variation in $\upgamma_{\rom{1}}$ to Type~II effects. An explanation is to be sought from a different dynamical effect. We thus postulate that a third critical solution, possibly another part of the Type~I attractor which is a family of metastable soliton stars, as a more reasonable explanation of our results.
\par A comparison with a study on Yang-Mills fields is also appropriate~\cite{nothingnew}. There, the authors consider the critical behavior of a two-parameter Yang-Mills kink, which like ours in different parts of their configuration space (depicted in their Fig. 4) variably exhibits Type~I and Type~II collapse or asymptotic dispersal. Their three boundaries correspond to the expected Type~I and Type~II critical solutions, as well as a third class of static colored black holes along the Type~I/Type~II interface. The authors entertain the thought of how a two-parameter massive field might exhibit similar behavior to their results, with the caveat that no-hair theorems dictate the nonexistence of static solutions that would be analogous to their Type~I/Type~II boundary.
\par Our paper does not exactly coincide with their work nor their analysis, since it concerns two fields that, near the triple point, are both simultaneously critical by themselves. For the reason, too, that our code was designed to work at such precision with mesh refinement as to ascertain the correct scaling, echoing, periodicity, and the like to numerical precision to solidify our claim of competing critical effects coupled only by the gravitational interaction between two separate fields, our method is not able to probe the post-collapse behavior. However, we can report the mass and time-to-collapse behavior near our Type~I/Type~II interface goes as expected, with the former decreasing towards the triple point and the latter increasing.
\par We close with a final mention of a curious behavior we have encountered. Between the triple point and the pure massive field regime in the phase portrait of scalar amplitudes, we find behavior a great deal more elaborate than that found by tracing out the Type~II section of the boundary. This vexing situation is shown in Fig.~\ref{fig:largephase}. The Choptuon being such a strong attractor likely explains why this effect was not seen along the Type~II section of the boundary. The influence of a hypothetical third critical solution, whose properties would seem necessarily similar to -- but not precisely the same as -- the metastable star, might be responsible for this interesting behavior. Another explanation could attribute the effect merely to a change in the dynamical behavior of the Type~I attractor: in this case, Fig.~\ref{fig:largephase} suggests that this evolution in itself is a rather involved phenomenon.

\section{Conclusion}
We have put the claims advanced in our previous paper~\cite{us} on more quantitative grounds by numerically investigating dominant perturbative exponents across the collapse/dispersal boundary of Fig.~\ref{fig:combinedcolor}, employing two separate methods as cross-checks to ensure consistency. We justified our categorization for the reason of the drastically different dominant exponents and time scales at play across the various sections of the dispersal/collapse boundary. Furthermore, our analysis of the varying dominant exponent along the Type~I section of the boundary suggests the existence of an emergent third critical solution, as we suggested following an alternative possibility considered by Gundlach et al.~\cite{panic}.
\par A large portion of the numerical region between the triple point and the regime of the pure massive field has proven to be numerically intractable using our current methods. While a third critical solution with influence near that of the Type~I solution could explain this difficult behavior, the lack of quantitive surety leaves much unexplained absent results by a satisfactory alternative approach.
\par Qualification, too, of the nature of the jumps we observe in the Type~I collapse times may warrant further investigation, clarifying whether such jumps are seen close to the triple point at greater precisions, and what determines the height of the jumps and the lengths of their associated ridges. For the reason that the slopes of the ridges reflect a change in the dominant perturbative exponent, we can wonder whether structures in the numerically intractable region exhibit features associated with such changes as well, since according to our hypothesis they share a common origin.
%\vfill\null
%\columnbreak
\begin{acknowledgments}
We gratefully acknowledge the technical support and computational resources of the Center for High Performance Computing at the University of Utah.
\end{acknowledgments}

\bibliography{ms}


\clearpage

\begin{figure}[htbp]
\includegraphicsdpi{\smalldpi}{width=0.95\linewidth}{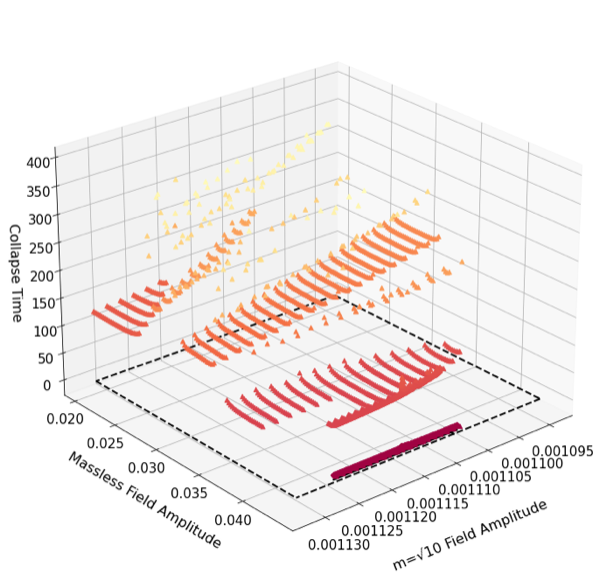}
   \caption{\label{fig:largephasetime} 3d scatterplot of the collapse time for the same region as depicted in Fig.~\ref{fig:largephase}. As in Fig.~\ref{fig:combinedcolortime}, the dashed line surrounds the sampled region. The various Type~I ``fingers'' have differing slopes, corresponding to distinct critical evolutions.}
\end{figure}

\begin{figure*}[!hb]
\centering
\includegraphicsdpi{\smalldpi}{width=0.9\linewidth}{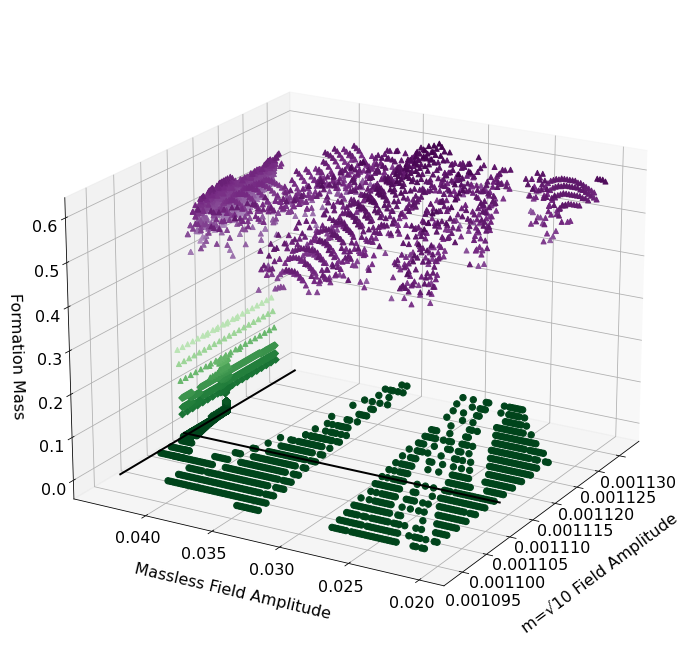}
   \caption{\label{fig:largephase} A broader view of the parameter space depicted in Fig.~\ref{fig:combinedcolor}. The classification used here is the same. The vertical and horizontal solid lines again reflect the but-for critical amplitudes of the appropriate fields. Behavior more intricate than that of Fig.~\ref{fig:combinedcolor} is observed, featuring ``fingers'' of dispersal with numerically difficult boundaries. Qualitatively and quantitatively these effects numerically persist at higher resolutions, although we cannot rule out that some of what is shown here may be numerical artifacts.}
\end{figure*}

\end{document}